\documentclass[aps,prl,superscriptaddress,floatfix,preprint]{revtex4-1}\usepackage{amsmath}
\usepackage{amsfonts}
\usepackage{amssymb}
\usepackage{epsfig}

\usepackage{graphicx}
\usepackage{color}
\usepackage{siunitx}

\begin{document}

\title{Josephson radiation from gapless Andreev bound states in HgTe-based topological junctions}

%\author{R. S. Deacon$^{1,2,\dagger}$, J. Wiedenmann$^{3,\dagger}$, E. Bocquillon$^{3,\dagger,\ast}$, F. Dom\'inguez$^{4}$,\\
%T. M. Klapwijk$^{5}$, P. Leubner$^{3}$, C. Br\"une$^{3}$, E. M. Hankiewicz$^{4}$,\\
%S. Tarucha$^{2,6}$, K. Ishibashi$^{2,3}$,  H. Buhmann$^{3}$, L. W. Molenkamp$^{3}$\\
%\\
%\normalsize{$^1$ Advanced Device Laboratory, RIKEN}\\ 
%\normalsize{2-1 Hirosawa, Wako-shi, Saitama, 351-0198, Japan}\\
%\normalsize{$^2$ Center for Emergent Matter Science, RIKEN}\\ 
%\normalsize{2-1 Hirosawa, Wako-shi, Saitama, 351-0198, Japan}\\
%\normalsize{$^{3}$ Physikalisches Institut (EP3), Universit\"at W\"urzburg}\\ 
%\normalsize{Am Hubland, D-97074 W\"urzburg, Germany}\\
%\normalsize{$^{4}$ Institut f\"ur Theoretische Physik (TP4), Universit\"at W\"urzburg}\\ 
%\normalsize{Am Hubland, D-97074 W\"urzburg, Germany}\\
%\normalsize{$^5$ Kavli Institute of Technology, Faculty of Applied Sciences,}\\
%\normalsize{Delft University of Technology}\\
%\normalsize{Lorentzweg 1, 2628 CJ Delft, The Netherlands}\\
%\normalsize{Department of Applied Physics, University of Tokyo}\\ 
%\normalsize{7-3-1 Hongo, Bunkyo-ku, Tokyo, 113-8656, Japan}\\
%\normalsize{$^\dagger$ All three authors contributed equally to this work.}\\
%\normalsize{$^\ast$ erwann.bocquillon@physik.uni-wuerzburg.de}}

\author{R.S.~Deacon}
\thanks{All three authors contributed equally to this work, email: erwann.bocquillon@physik.uni-wuerzburg.de}
\affiliation{Advanced Device Laboratory, RIKEN, 2-1 Hirosawa, Wako-shi, Saitama, 351-0198, Japan}
\affiliation{Center for Emergent Matter Science, RIKEN, 2-1 Hirosawa, Wako-shi, Saitama, 351-0198, Japan}

\author{J.~Wiedenmann}
\thanks{All three authors contributed equally to this work, email: erwann.bocquillon@physik.uni-wuerzburg.de}
\affiliation{Physikalisches Institut (EP3), Universit\"at W\"urzburg, Am Hubland, D-97074 W\"urzburg, Germany}

\author{E.~Bocquillon}
\thanks{All three authors contributed equally to this work, email: erwann.bocquillon@physik.uni-wuerzburg.de}
\affiliation{Physikalisches Institut (EP3), Universit\"at W\"urzburg, Am Hubland, D-97074 W\"urzburg, Germany}

\author{F.~Dominguez}
\affiliation{Institut f\"ur theoretische Physik (TP4), Universit\"at W\"urzburg, Am Hubland, D-97074 W\"urzburg, Germany}

\author{T.M.~Klapwijk}
\affiliation{Kavli Institute of Nanoscience, Faculty of Applied Sciences, Delft University of Technology, Lorentzweg 1, 2628 CJ Delft, The Netherlands}
%\affiliation{Laboratory for Quantum Limited Devices, Physics Department, Moscow State Pedagogical University, 29 Malaya Pirogovskaya St. Moscow 119992, Russia}

\author{P.~Leubner}
\affiliation{Physikalisches Institut (EP3), Universit\"at W\"urzburg, Am Hubland, D-97074 W\"urzburg, Germany}
%
%\author{C.~Ames}
%\affiliation{Physikalisches Institut (EP3), Universit\"at W\"urzburg, Am Hubland, D-97074 W\"urzburg, Germany}

\author{C.~Br\"une}
\affiliation{Physikalisches Institut (EP3), Universit\"at W\"urzburg, Am Hubland, D-97074 W\"urzburg, Germany}

\author{E.M.~Hankiewicz}
\affiliation{Institut f\"ur theoretische Physik (TP4), Universit\"at W\"urzburg, Am Hubland, D-97074 W\"urzburg, Germany}

\author{S.~Tarucha}
\affiliation{Center for Emergent Matter Science, RIKEN, 2-1 Hirosawa, Wako-shi, Saitama, 351-0198, Japan}
\affiliation{Department of Applied Physics, University of Tokyo, 7-3-1 Hongo, Bunkyo-ku, Tokyo, 113-8656, Japan}

\author{K.~Ishibashi}
\affiliation{Advanced Device Laboratory, RIKEN, 2-1 Hirosawa, Wako-shi, Saitama, 351-0198, Japan}
\affiliation{Center for Emergent Matter Science, RIKEN, 2-1 Hirosawa, Wako-shi, Saitama, 351-0198, Japan}

\author{H.~Buhmann}
\affiliation{Physikalisches Institut (EP3), Universit\"at W\"urzburg, Am Hubland, D-97074 W\"urzburg, Germany}

\author{L.W.~Molenkamp}
\affiliation{Physikalisches Institut (EP3), Universit\"at W\"urzburg, Am Hubland, D-97074 W\"urzburg, Germany}

\begin{abstract}
Frequency analysis of the rf emission of oscillating Josephson supercurrent is a powerful passive way of probing properties of topological Josephson junctions. In particular, measurements of the Josephson emission enables to detect the expected presence of topological gapless Andreev bound states that give rise to emission at half the Josephson frequency $f_J$, rather than conventional emission at $f_J$. Here we report direct measurement of rf emission spectra on Josephson junctions made of HgTe-based gate-tunable topological weak links. The emission spectra exhibit a clear signal at half the Josephson frequency $f_{\rm J}/2$. The linewidths of emission lines indicate a coherence time of $0.3-\SI{4}{ns}$ for the $f_{\rm J}/2$ line, much shorter than for the $f_{\rm J}$ line ($3-\SI{4}{ns}$). These observations strongly point towards the presence of topological gapless Andreev bound states, and pave the way for a future HgTe-based platform for topological quantum computation.
\end{abstract}

\maketitle

\noindent

\newpage 

%As a consequence of the Josephson equations, a strong relation exists between the voltage $V$ measured across a Josephson junction and the characteristic frequency $f_{\rm J}=\frac{2eV}{h}$ of the current flowing in it. 'Listening' to the rf signal radiated by the circulating supercurrent is a powerful passive way of probing its properties. In Josephson junctions based on topological weak links, it enables in particular to identify the contribution of gapless Andreev bound states that directly manifest themselves as an emission line at half the Josephson frequency $f_{\rm J}/2$. We report direct measurement of rf emission spectra on Josephson junctions made of HgTe-based gate-tunable topological weak links. The emission spectra exhibit a clear signal at $f_{\rm J}/2$ and are consistent with the presence of topological quantum spin Hall edge states. Besides, the widths of emission lines indicates a coherence time of $0.3-\SI{4}{ns}$ for the $f_{\rm J}/2$ line, much shorter than for the $f_{\rm J}$ line ($3-\SI{4}{ns}$).

\begin{figure}[h!]
\centerline{\includegraphics[width=0.5\textwidth]{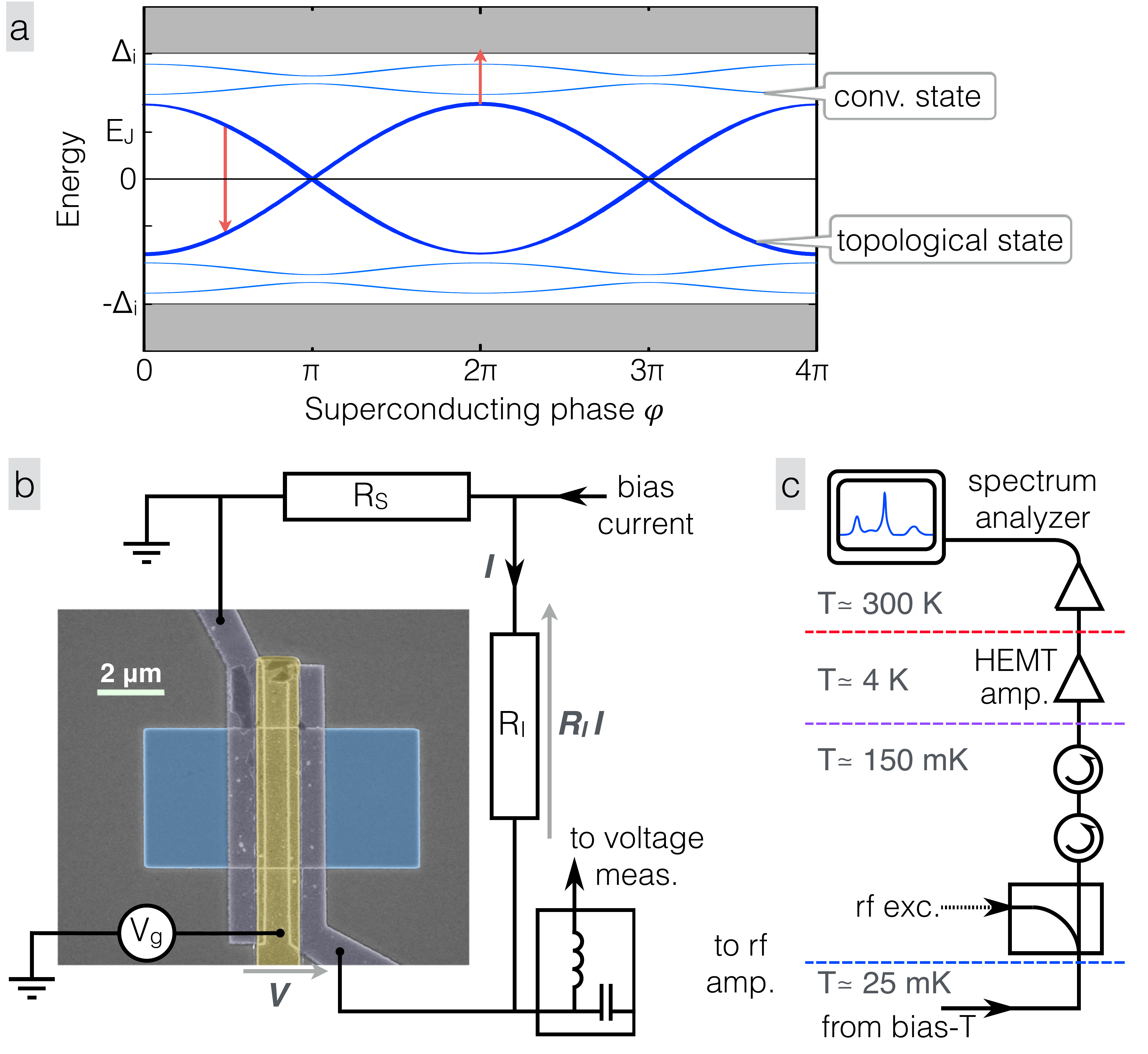}}
\caption{{\bf Voltage bias and measurement setups -} a) The spectrum of Andreev bound states in a topological Josephson junction hosts a $4\pi$-periodic gapless Andreev doublet (dark blue) around zero-energy, and other gapped Andreev modes (bulk or edge modes, depicted in light blue). Coupling to the continuum (grey area), other Andreev states, and possible relaxation mechanisms are pictured as red arrows. The exact spectrum is not known, and depends on parameters such as length of the junction and Fermi energy. b) On the colored SEM picture, the mesa is visible in blue, the Al leads in violet, and the gate is indicated in yellow. A shunt resistance $R_S$ enables a stable voltage bias across the junction. The measurement of the voltage across the junction and across $R_I$ directly yields $V$ and $I$. The rf signal is coupled to the amplification scheme via a bias-T. c) The rf signal is amplified by a cryogenic HEMT amplifier (with additional room-temperature amplification). A directional coupler can be used to send an rf excitation signal (backwards) for characterization purposes (see SI).} \label{Fig:1Setups}
\end{figure}

%\subsection{Introduction}
In recent years, schemes for fault-tolerant quantum computation have been theoretically developed on the premises of non-abelian particle statistics \cite{Kitaev2003}. Such statistics can arise in condensed matter systems for so-called Majorana quasiparticles, that may be braided around one another to execute quantum information protocols. Majorana zero-modes can be conveniently engineered by inducing $p$-wave superconductivity in a two-dimensional topological insulator \cite{Alicea2012,Beenakker2013} (2D TI).
Coupling the topological edge channels of a 2D TI to a nearby conventional $s$-wave superconductor leads to the appearance of an induced $p$-wave superconducting phase \cite{Fu2009}. In a topological Josephson junction, a doublet of $p$-wave Andreev bound states is predicted to have a topologically protected level crossing for a superconducting phase difference $\varphi=\pi,3\pi,...$ across the junction. Such states can in principle be detected via the resulting energy dispersion that is $4\pi$-periodic in $\varphi$, with, in the simplest case of a short junction, $E=E_{\rm J}\cos\varphi/2$ \cite{Kwon2003,Fu2009}. However in the thermodynamic limit of a time-independent phase $\varphi$, the current is $2\pi$-periodic as only the lower branch at $E\leq0$ is populated. Experiments relying on out-of-equilibrium dynamics in the GHz range are thus useful to provide evidence for the existence of gapless $4\pi$-periodic Andreev bound states on time scales shorter than equilibration time. Equilibration occurs through various relaxation processes such as coupling to the continuum, to other Andreev bound states, or quasiparticle poisoning \cite{Kwon2003, Badiane2011,Pikulin2012,SanJose2012}, see Fig.\ref{Fig:1Setups}a. On such short time scales, Josephson emission at half the Josephson frequency $f_{\rm J}/2$ is then predicted \cite{Fu2009,Badiane2011,SanJose2012}.
%%%%%%%%

In previous works \cite{Wiedenmann2016,Bocquillon2016}, we reported a doubling of the Shapiro step size ($hf/e$) in Josephson weak links based on thick strained HgTe layers (3D TI) and HgTe quantum wells (2D TI) which clearly indicates the presence of a $4\pi$-periodic component in the supercurrent. Though experimentally easily accessible, detailed interpretation of such experiments is hindered by the strongly non-linear nature of the Josephson response to an rf excitation. In contrast, Josephson emission under a dc voltage bias provides a passive and direct probe of supercurrents in topological junctions, but the radiated power is low and difficult to measure. Moreover, the linewidths of the emission lines reflect the lifetime of the Andreev bound states. In this article, we report on the study of Josephson emission in a range from 2 to \SI{10}{\giga\hertz} using cryogenic microwave measurements. Besides conventional emission at $f_{\rm J}$ and $2f_{\rm J}$, we observe clear emission at $f_{\rm J}/2$  in Josephson junctions based on inverted HgTe quantum wells, which are 2D TIs and exhibit the quantum spin Hall effect \cite{Konig2007}. These emission measurements provide very direct evidence of the presence of a $4\pi$-periodic supercurrent. Additionally, the coherence time of the unconventional emission line at $f_{\rm J}/2$ is observed to be up to an order of magnitude shorter than that at $f_{\rm J}$, indicating its sensitivity to relaxation processes. This set of experimental signatures is attributed to the presence of gapless Andreev bound states. In a reference experiment, a non-topological HgTe-based superconducting weak link exhibits only conventional emission at $f_{\rm J}$.

%The main text of this article focuses on the study of Josephson junctions made of HgTe quantum wells, similar results on 3D topological insulators made of thicker strained layers of typically \SI{70}{\nano\meter} are presented in the SI.

%\subsection{Samples and measurement setup}
Due to a band inversion, HgTe quantum wells become 2D topological insulators for a thickness larger than a critical thickness $d_c\simeq \SI{6.3}{\nano\meter}$. In this regime, they exhibit a pair of counter-propagating edge channels when tuned into the band gap (quantum spin Hall effect \cite{Konig2007}). In contrast, thinner HgTe wells have a trivial band structure. We analyze the different behaviors of measurements on a non-inverted trivial narrow quantum well (thickness $d\simeq \SI{5}{\nano\meter}$) with a conventional behavior, and a topological quantum well ($d\simeq \SI{8}{\nano\meter}$) that exhibits anomalous emission features. A false-colored SEM picture of a device is shown in Fig.\ref{Fig:1Setups}b. The HgTe heterostructure is shaped into a rectangular mesa, and contacted via two superconducting Al leads \SI{600}{\nano\meter} apart, to form a Josephson junction. The application of a voltage $V_g$ on a top gate enables tuning of the electron density in the weak link, to access the quantum spin Hall regime with edge channel conduction, as well as transport in the conduction or valence band.

In this experiment we directly measure with a spectrum analyzer the Josephson radiation emitted from HgTe-based Josephson junctions. For a given dc voltage $V$, the phase difference across the Josephson junction evolves with time, and the resulting oscillatory current can be measured and analyzed using rf techniques. While early measurements of Josephson emission used narrow-band resonant cavities \cite{Yanson1965,Pedersen1976}, direct wide-band measurements of emission spectra are nowadays accessible via microwave cryogenic amplifiers \cite{Schoelkopf1995}. To this end, the junction is connected to a coaxial line and decoupled from the dc measurement line via a bias-T (see Fig.\ref{Fig:1Setups}b and Fig.\ref{Fig:1Setups}c, and SI for details). The rf signal is then amplified at both cryogenic and room temperatures before being measured with a spectrum analyzer. The commercial rf components used in the readout line limit the frequency range of detection to 2-\SI{10}{\giga\hertz}.
An essential requirement to successfully perform such measurements is a stable bias. Under current bias, instabilities and hysteretic behavior may occur at low voltages \cite{Wiedenmann2016,Bocquillon2016}. We therefore employ a small resistive shunt $R_S$ (between 1 and \SI{50}{\ohm}) to enable a stable voltage bias (though residual switching below a few microvolts is sometimes seen), while a small resistance $R_I$ in series with the junction yields a measurement of the current $I$ through the junction (Fig.\ref{Fig:1Setups}c). 

%\subsection{Emission spectra and frequency dependence}

\begin{figure*}[h!]
\centerline{\includegraphics[width=1\textwidth]{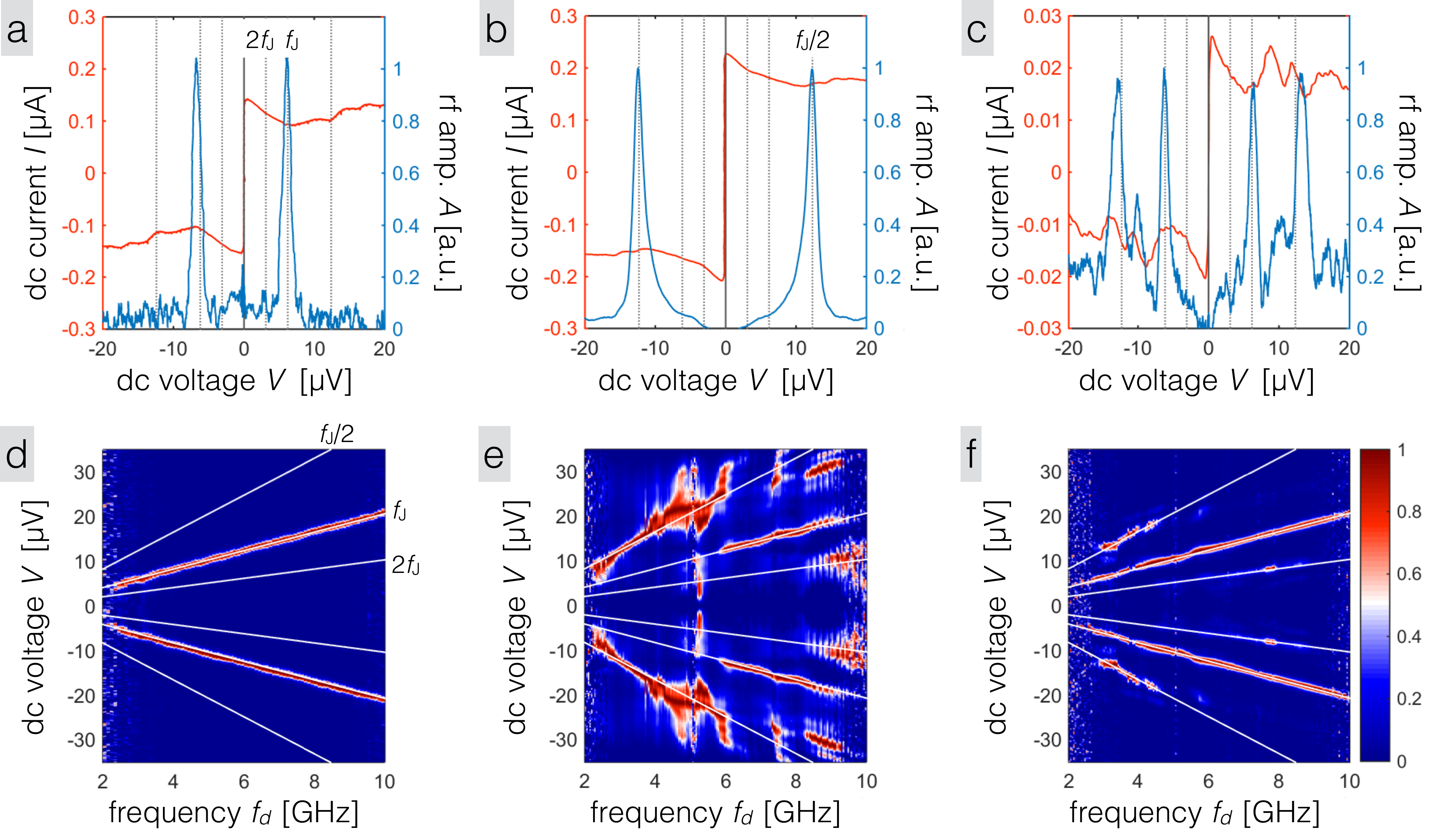}}
\caption{{\bf Emission spectra -} a,b,c) $I$-$V$ curves (red lines) and emission spectrum (blue lines) for a non-topological weak link (a) and a quantum spin Hall weak link (b at $V_g=\SI{-0.55}{\volt}$ and c at $V_g=\SI{-1.4}{\volt}$). The radiation is collected at fixed detection frequency $f_d\simeq\SI{3}{\giga\hertz}$ when sweeping the bias. Grey dashed lines indicate the expected values of $2f_{\rm J}$ (innermost), $f_{\rm J}$ and $f_{\rm J}/2$ (outermost lines). d,e,f) Two-dimensional plot of the power emitted by the devices, as function of voltage $V$ and detection frequency $f_d$ (d in the non-topological weak link, e and in the topological weak link for respectively $V_g=\SI{-0.55}{\volt}$ and $V_g=\SI{-1.4}{\volt}$). White lines indicate the expected resonance lines at $2f_{\rm J}$ (innermost), $f_{\rm J}$ and $f_{\rm J}/2$ (outermost). For better visibility, the data is normalized to its maximum for each frequency.} \label{Fig:2FreqDep}
\end{figure*}

In the absence of any drive on the junction, a background noise is observed, probably originating from black body radiation and parasitic stray noise from the environment. It is taken as a reference and subtracted from all measurements to isolate the contribution of the junction. When the junction is biased and a finite voltage $V$ develops, the contribution of the junction appears. In Fig.\ref{Fig:2FreqDep} (a,b,c), the amplitude $A$ of the rf signal collected at the fixed detection frequency of $f_d=\SI{3}{\giga\hertz}$ (in an $\SI{8}{\mega\hertz}$ bandwidth) is plotted as a blue line, as function of the voltage $V$ across the junction. $I$-$V$ curves and emission spectra are taken for topologically trivial and nontrivial samples, over large range of gate voltages. Three $I$-$V$ curves are presented (as red lines), showing that a stable voltage bias is obtained down to a few microvolts. In the case of a non-topological well (Fig.\ref{Fig:2FreqDep}a), we observe a very clear peak when the Josephson radiation matches the detection frequency $f_d=f_{\rm J}=\frac{2eV}{h}$, regardless of the position of the Fermi energy. This agrees with early observations on (non-topological) microbridges \cite{Yanson1965,Pedersen1976}. In strong contrast, the topological Josephson junction unveils a new and strong feature at half the Josephson frequency $f_{\rm J}/2$ (sometimes concomitant with emission at $f_{\rm J}$). This observation, illustrated for two different electron densities in Fig.\ref{Fig:2FreqDep}b and \ref{Fig:2FreqDep}c, is a direct manifestation of the presence of a $4\pi$-periodic supercurrent flowing in our topological junctions \cite{Badiane2011} and constitutes our main finding. In Fig.\ref{Fig:2FreqDep}b, measured in the vicinity of the quantum spin Hall regime, only the line at  $f_{\rm J}/2$ is visible. In Fig.\ref{Fig:2FreqDep}c, measured towards the valence band ($p$ conduction regime), both lines at  $f_{\rm J}$ and  $f_{\rm J}/2$ are observed. We review in detail the effect of the position of the Fermi level further in the article.

When the detection frequency is swept, one can verify that the emission lines follow the linear relation $f_{\rm J}=\frac{2eV}{h}$. In the weak link with a trivial band structure (Fig.\ref{Fig:2FreqDep}d), the conventional $2\pi$-periodic line is visible over the range $2-\SI{10}{\giga\hertz}$ (for each value of $f_d$, the reference at $I=0$ is subtracted, and the data is normalized to its maximum to correct for frequency-dependent coupling and amplification).
In the topological device and for $V_g=\SI{-0.55}{\volt}$ (Fig.\ref{Fig:2FreqDep}e), the colormap shows that the emission is entirely dominated by the $4\pi$-periodic supercurrent below $f=\SI{5.5}{\giga\hertz}$, before the conventional line is recovered. At higher frequencies, the emission spectrum is influenced by resonant modes within the electromagnetic environment. These can be easily identified by a characterization of the electromagnetic environment of the junction (see SI).
When $V_g\simeq\SI{-1.4}{\volt}$ (Fig.\ref{Fig:2FreqDep}f), the colormap reveals that the $4\pi$-periodic component at $f_{\rm J}/2$ is visible only up to $f_d\simeq\SI{4.5}{\giga\hertz}$, while the conventional emission line at $f_{\rm J}$ is seen in the whole range of frequencies. 

\begin{figure*}[h!]
\centerline{\includegraphics[width=1\textwidth]{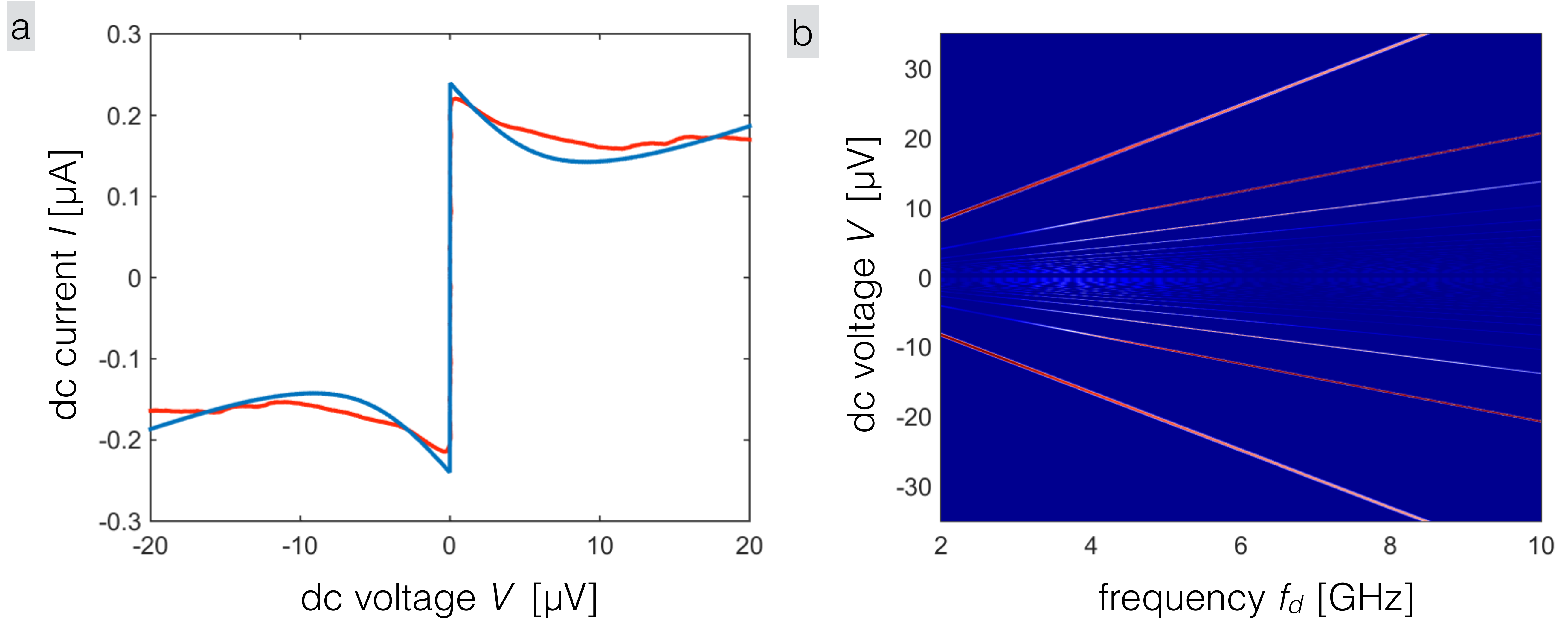}}
\caption{{\bf RSJ simulations -} a) Simulated $I$-$V$ curve (blue) compared with measured data at $V_g=\SI{-0.55}{\volt}$. The simulations are performed for $R_I=R_S=\SI{24}{\ohm},\, R_n=\SI{130}{\ohm},\, I_{4\pi}=\SI{100}{\nano\ampere}$, and $I_c=\SI{240}{\nano\ampere}$. b) Simulated Fourier transform of the voltage $V$ in the junction, as function of detection frequency $f_d$ and voltage $V$, for the same simulation parameters as in a). A good qualitative agreement is found with Fig.\ref{Fig:2FreqDep}e. Especially, the predominance of the emission at $f_{\rm J}/2$ for low voltages (below \SI{12}{\micro\volt}) is well reproduced.} \label{Fig:3Simus}
\end{figure*}

We now analyze the measurements of Fig.\ref{Fig:2FreqDep}e. The strong dominance of $4\pi$-periodic radiation observed in Fig.\ref{Fig:2FreqDep}e at low frequencies/voltages may at first sight be surprising, as conventional $2\pi$-periodic modes are also expected to contribute. To model the experimental data, we have performed numerical simulations, in the framework of a Resistively Shunted Junction (RSJ) model, modified to account for the shunt circuit and the $4\pi$-periodic component of the supercurrent (see SI). We compute successively the time-dependent voltage $V(t)$ and its Fourier transform to obtain the amplitudes of each frequency component. The non-linear response to the two time scales associated to the combination of $2\pi$-/$4\pi$-periodic contributions allow $4\pi$-periodic (resp. $2\pi$-) terms to be more visible for low (high) voltages. For currents such that $I_c\lesssim I\lesssim I_c+I_{4\pi}$ (with $I_c$ the total critical current, and $I_{4\pi}$ the amplitude of the $4\pi$-periodic contribution to $I_c$), the dynamics of the junction is highly non-linear, the $2\pi$-periodic component of the voltage is effectively suppressed, resulting in $4\pi$-periodic oscillating voltages\cite{Dominguez2012} and emission at $f_{\rm J}/2$. Consequently, for the corresponding low voltages, the junction is expected to emit mainly at $f_{\rm J}/2$. For higher biases, the dynamics of the system is ruled by a single time scale, and resembles that of a $2\pi$-periodic junction. Computations for increasing voltages $V$ and detection frequency $f_d$ yield a good qualitative agreement with the $I$-$V$ characteristic (Fig.\ref{Fig:3Simus}a) as well as the emission features (Fig.\ref{Fig:3Simus}b) for a contribution of $4\pi$-periodic modes amounting to around 40\% of the critical current, in agreement with previous estimates \cite{Bocquillon2016}.

\begin{figure}[h!]
\centerline{\includegraphics[width=0.5\textwidth]{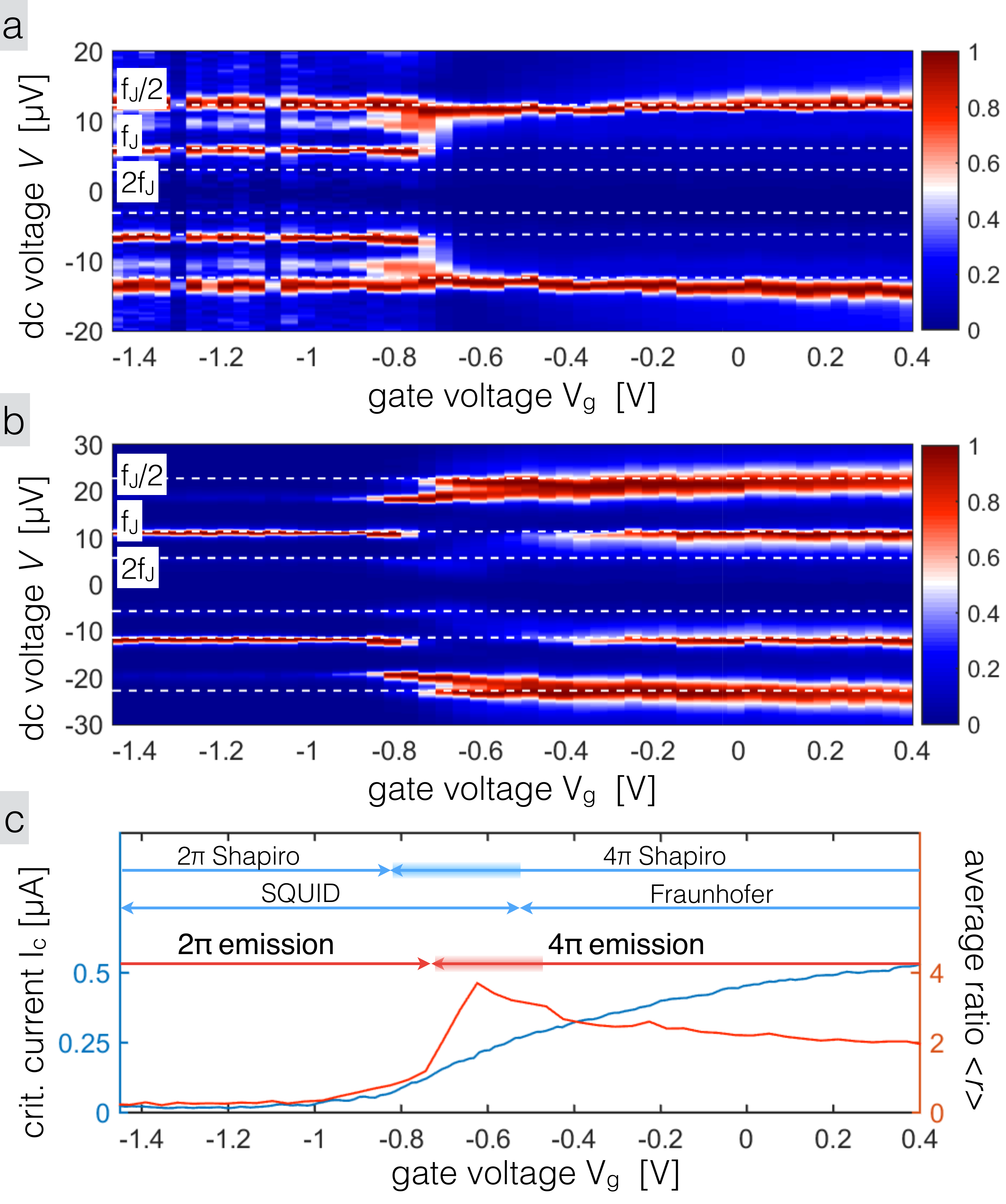}}
\caption{{\bf Gate Dependence -} a,b) 2D maps of the normalized amplitude $A$ as function of bias voltage $V$ and gate voltage $V_g$ for detection frequency $f_d=\SI{2.98}{\giga\hertz}$ and \SI{5.5}{\giga\hertz} respectively. c) The ratio of the intensity of the $4\pi$- to the $2\pi$-periodic amplitudes $\langle r\rangle=\langle \frac{A_{4\pi}}{A_{2\pi}}\rangle$ (averaged over frequency in 2-\SI{10}{\giga\hertz} range) (as a red line) and the critical current (as a blue line) are plotted as function of gate voltage $V_g$. After rescaling of the critical current measured in Ref.\cite{Bocquillon2016}, the emission features can be compared with previous observations on the Shapiro response.} \label{Fig:3GateDep}
\end{figure}

We now detail the dependence of the emitted power as a function of the gate voltage. In the non-topological device (see SI), we observe that the amplitude of the collected signal reflects the amplitude of the critical current, and verify that $A\propto I_c$ with a good agreement \cite{Likharev1986}. This confirms the conventional behavior of the device in the conduction and valence bands of the quantum well, as well as close to the gap. In Fig.\ref{Fig:3GateDep}a and \ref{Fig:3GateDep}b, we present two sets of measurements of the collected rf amplitude $A$ on the topological weak link, taken at a low ($f_d=\SI{2.98}{\giga\hertz}$) and high ($f_d=\SI{5.5}{\giga\hertz}$) frequency. We observe three clear regimes in the emitted power that correlate with the expected band structure. When the gate voltage is above $V_g>\SI{-0.4}{\volt}$, we observe that emission occurs for $f_d=f_{\rm J}/2$ at $f_d=\SI{2.98}{\giga\hertz}$ (Fig.\ref{Fig:3GateDep}a), and for both $f_{\rm J}$ and for $f_{\rm J}/2$ at high frequency of $f_d=\SI{5.5}{\giga\hertz}$ (Fig.\ref{Fig:3GateDep}b). These observations suggest transport in the conduction band of the quantum well, where gapless Andreev bound states have been seen to coexist with $n$-type conventional states, in agreement with previous observations and predictions \cite{Bocquillon2016,Dai2008}. When $\SI{-0.8}{\volt}<V_g<\SI{-0.4}{\volt}$, one observes almost exclusively emission at half the Josephson frequency $f_d=f_{\rm J}/2$. This voltage range corresponds to the quantum spin Hall regime where edge states are the dominant transport channel. This observation is thus in line with the topological origin of this anomalous spectral line. For $V_g<\SI{-0.8}{\volt}$, we observe Josephson radiation at both $f_{\rm J}$ and $f_{\rm J}/2$ which suggests the coexistence of weak gapless Andreev modes with bulk $p$-type conventional modes of the valence band. The overall gate voltage dependence is consistent with the expected band structure of a quantum spin Hall insulator. Finally, we compute from the measurement data the amplitude $A_{2\pi},A_{4\pi}$ extracted along the emission line at $f_{\rm J}$ and $f_{\rm J}/2$ respectively. For each gate voltage $V_g$, we calculate the ratio $r=\frac{A_{4\pi}}{A_{2\pi}}$ (which is independent of the frequency response of the amplification scheme). We plot this ratio $\langle r \rangle$ averaged over frequency $f_d$ as a red line in Fig.\ref{Fig:3GateDep}c and compare our results with information inferred from the Shapiro response in our previous work \cite{Bocquillon2016}. As expected from Fig.\ref{Fig:3GateDep}a and \ref{Fig:3GateDep}b, $\langle r \rangle$ shows a maximum around $V_g=\SI{-0.7}{\volt}$, where the conventional line is strongly suppressed. When correcting for density differences (offset on the gate voltage) and the strength of induced superconductivity (multiplication factor on the current), the critical current in the present experiment can be mapped to the one measured in Ref.\cite{Bocquillon2016}. We observe that a good agreement is found between the observation of edge transport (SQUID-like response to a magnetic field), the $4\pi$-periodic Shapiro response, and the emission at the half the Josephson frequency $f_{\rm J}/2$.

Beyond the direct detection of a $4\pi$-periodic supercurrent, the measurement of Josephson emission in these devices provides other new insights. First, we detect emission at $f_{\rm J}/2$ as low as $\SI{1.5}{\giga\hertz}$. If responsible for the $4\pi$-periodic supercurrent\cite{Billangeon2007,Dominguez2012}, potential Landau-Zener transitions would be activated for a voltage $V_{LZ}\ll\SI{6}{\micro\volt}$. This sets an upper bound on the existence of a residual avoided crossing $\delta\ll\sqrt{\frac{V_{LZ}E_J}{8\pi}}=\SI{5}{\micro\electronvolt}$ \cite{Pikulin2012,Virtanen2013}, and tend to rule this mechanism as origin for the $4\pi$-periodic emission. Furthermore, the linewidth of both emission lines can be examined. For conventional Josephson radiation, the linewidth is in principle related to fluctuations in the pair or quasiparticle currents \cite{Stephen1968,Dahm1969} or can be dominated by the noise in the environment \cite{Likharev1986}. In both the topological and trivial devices, the line at $f_{\rm J}$ exhibits a typical width of $\delta V_{2\pi}\simeq 0.5 - \SI{0.8}{\micro\volt}$, i.e. a coherence time $\tau_{2\pi}=\frac{h}{2e\delta V}\simeq 3 - \SI{4}{\nano\second}$ (when the anomalous emission at $f_{\rm J}/2$ is absent). This width is consistent with Shapiro steps\cite{Bocquillon2016} being observable down to typically \SI{0.5}{\giga\hertz}. In contrast, the linewidth at $f_{\rm J}/2$ can additionally reflect relaxation mechanisms such as ionization to the continuum \cite{Badiane2011,Badiane2013} or parity relaxation mechanisms. We observe that the linewidth at $f_{\rm J}/2$ varies more strongly and obtain typical widths in the range $\delta V_{4\pi}\simeq 0.5 - \SI{8}{\micro\volt}$ yielding a shorter coherence time $\tau_{4\pi}\simeq 0.3-\SI{4}{\nano\second}$. As visible in Fig.\ref{Fig:3GateDep}a and \ref{Fig:3GateDep}b, the linewidth increases when the gate voltage is driven deeper in the conduction band. While it is difficult to establish clear trends due to the complex high frequency response, the linewidth also seem to increase when the frequency (or equivalently voltage bias) is increased. Both observations may signal a decrease of lifetime when the $4\pi$-periodic modes are coupled to an increasing number of $2\pi$-periodic modes or to the continuum via ionization processes.

To conclude, we here demonstrate the emission of topological junctions at half the Josephson frequency $f_{\rm J}/2$. Our results tend to confirm that the observed $4\pi$-periodic response results from gapless Andreev bound states. Moreover, they indicate the absence of Landau-Zener activation, and provides additional information on the lifetime of these gapless Andreev bound states.

{\bf Acknowledgments:}
We gratefully acknowledge insightful discussions with B.~Trauzettel, \c C.~Girit, M.~Hofheinz, F.~Portier, H.~Pothier, L.~Fu, R.~Aguado, C.~Lobb, M.~Houzet, and J.S.~Meyer.
This work is supported by the German Research Foundation (Leibniz Program, SFB1170 Tocotronics) and the Elitenetzwerk Bayern program €œTopologische Isolatoren€. R.S.D. acknowledges support from Grants-in-Aid for Young Scientists B (No. 26790008) and Grants-in-Aid for Scientific Research (No. 16H02204). T.M.K. is financially supported by the European Research Council Advanced grant No.339306 (METIQUM) and by the Ministry of Education and Science of the Russian Federation under Contract No.14.B25.31.007. S.T. acknowledges financial support from Grants-in-Aid for Scientific Research S (No. 26220710), MEXT and ImPACT Program of Council for Science, Technology and Innovation. E.B., T.M.K. and L.W.M. gratefully thank the Alexander von Humboldt foundation for its support.

\clearpage
\begin{center}
{\bf \center \large Josephson radiation from gapless Andreev bound states} \\
{\bf \large in HgTe-based topological junctions}\\
{\bf \large --}\\
{\bf \large Supplementary Information}
\end{center}

\vspace{2\baselineskip}

\section{Detailed description of the setup}
We detail here the setup used to measure the Josephson radiation of our devices. Results were collected in two different dilution refrigerator systems, which differed in the design of filtering and sample enclosures. In both cases, a PCB with a coplanar transmission line collects the radiation emitted from the device and couples through an SMA launcher to the rf measurement setup depicted in Fig.1c. The rf coupling line is first decoupled from the dc line via a bias tee. The signal line is amplified by a cryogenic HEMT amplifier (+\SI{39}{\decibel}) and two room-temperature amplifiers (each +\SI{30}{\decibel}) before reaching a spectrum analyzer (Keysight EXA N9010A). The commercial rf components used in the readout line limit the frequency range of detection to 2-\SI{10}{\giga\hertz}. Additionally, a directional coupler allows the input of external rf drive through the \SI{-20}{\decibel} port to probe the electromagnetic environment of the junction and perform Shapiro step measurements \cite{Wiedenmann2016,Bocquillon2016}.

\begin{figure*}[h!]
\centerline{\includegraphics[width=0.9\textwidth]{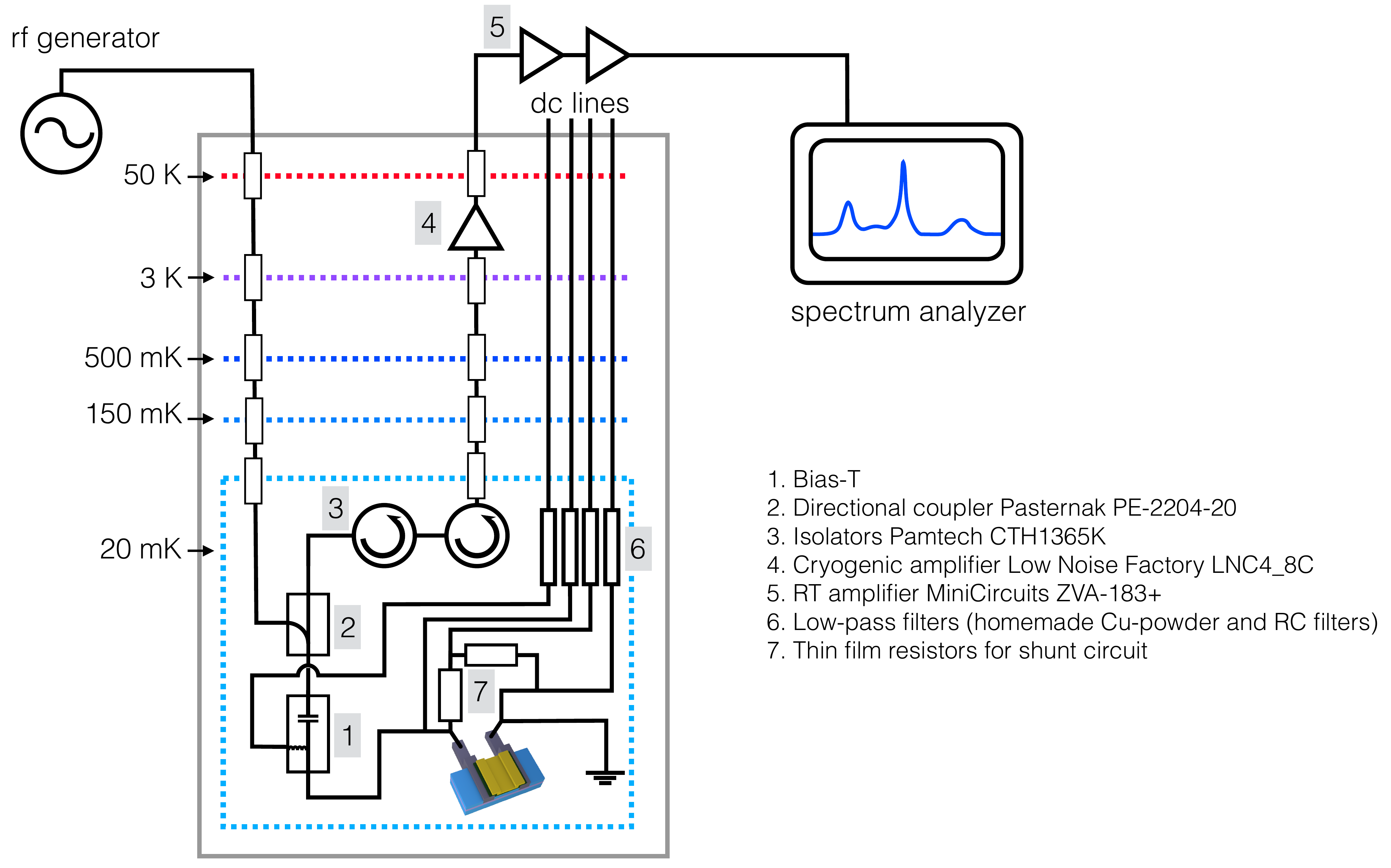}}
\caption{{\bf Detailed measurement setup -} The sample, thin film resistors, bias-tee and directional coupler sit near the mixing chamber of a cryo-free dilution refrigerator, at a base temperature of \SI{20}{\milli\kelvin}. Cryogenic and room-temperature amplification enable measurement of powers as low as \SI{0.1}{\femto\watt} in 2-\SI{10}{\giga\hertz} range.} \label{Fig:SOMRSJsetups}
\end{figure*}

The shunt and measurement resistors are thin metal film chip resistors placed in close proximity to the device to minimize stray inductance \cite{Chauvin2006}. In a typical measurement the detection frequency is fixed with a wide resolution band width of \SI{8}{\mega\hertz} while the $I$-$V$ characteristic of the junction is swept using a triangular waveform generator (Keysight 33250A). The junction bias and current are amplified (differential voltage amplifiers Femto DLPVA or similar) and measured with a digitizer (Rohde \& Schwarz RTO 1022) which is synchronized along with the spectrum analyzer to the sweeps of the junction current. Both the spectral data and the $I$-$V$ are averaged over several hundred repetitions to improve the signal to noise ratio at frequencies with weak signal. Such a setup enables a sensitivity of approximately \SI{0.1}{\femto\watt} throughout a 2-\SI{10}{\giga\hertz} range.

\section{Dynamics and RSJ simulations}

\subsection{RSJ model in the presence of $2\pi$- and $4\pi$-periodic supercurrents}

\paragraph{RSJ equation} In this section we present the results of numerical simulations carried out in the framework of the RSJ model. The junction is modeled by its current phase relation (CPR) $I_S(\varphi)$ together with a resistive shunt $R_n$ (see Fig.\ref{Fig:SOMRSJsetups}a), and a current bias $I$ is applied. When combined with the Josephson equation, one then readily obtains a first order differential equation\cite{Stewart1968,McCumber1968} on the superconducting phase difference $\varphi(t)$ that can be solved and consequently yields the voltage $V(t)=\frac{\hbar}{2e}\frac{d\varphi}{dt}$:
\begin{eqnarray}
\frac{\hbar}{2e R_n}\frac{d\varphi}{dt}+I_S(\varphi)=I.
\label{Eq:RSJ}
\end{eqnarray}

In previous works\cite{Dominguez2012,Wiedenmann2016}, we have studied an extended RSJ model which simulated the effect a $4\pi$-periodic contribution in the CPR (simply written as $I_S(\varphi)=I_{2\pi}\sin\varphi + I_{4\pi}\sin\varphi/2$). This differential equation is highly non-linear, and it is of importance in determining the parameter space for which a 4$\pi$-periodic contribution can be observed even in the case where it is accompanied by a larger $2\pi$-periodic contribution.

\paragraph{Case of a $2\pi$-periodic supercurrent}

\begin{figure*}[h!]
\centerline{\includegraphics[width=\textwidth]{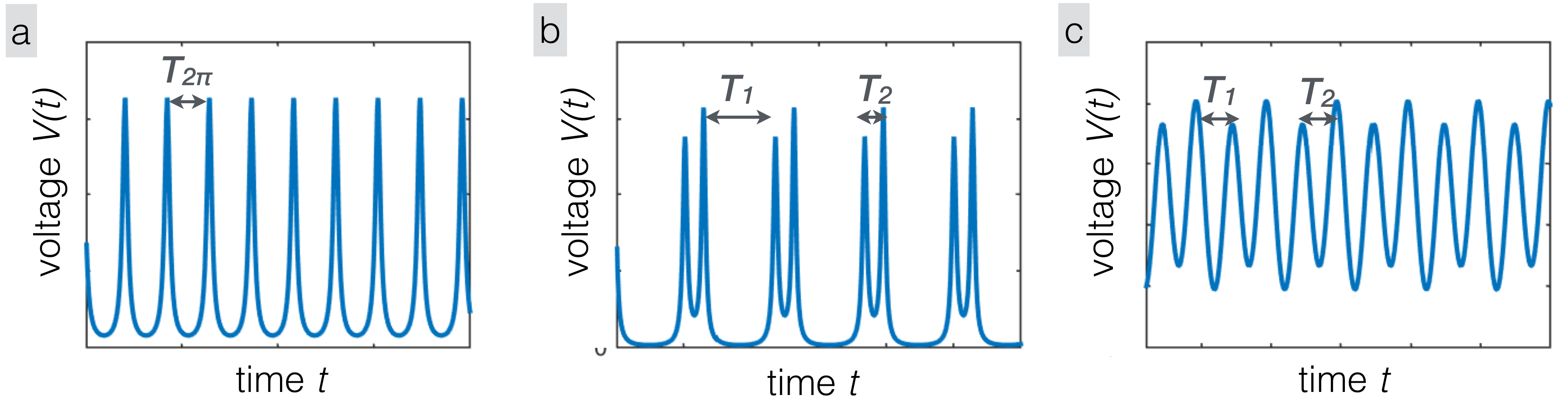}}
\caption{{\bf Time-dependent voltage -} a) For a $2\pi$-periodic CPR, $V(t)$ oscillates in time with a period $T_{2\pi}\propto \langle V\rangle^{-1}$. For $I\simeq I_{2\pi}, \langle V\rangle\simeq 0$ and $V(t)$ is highly non sinusoidal. Simulations computed for $I=1.05 I_{2\pi}$.  b) When a $4\pi$-periodic current is added to the CPR, it dominates the dynamics at low voltages, and $T_1\gg T_2$. In these conditions, the voltage is non-sinusoidal. Simulations computed for $I=1.05 I_c, I_{4\pi}=0.2 I_{2\pi}$. c) A mostly $2\pi$-periodic dynamics is recovered at higher voltages, and $T_1\simeq T_2\simeq T_{2\pi}$. The voltage becomes closer to a $2\pi$-periodic sine wave. Simulations computed for $I=8 I_c, I_{4\pi}=0.2 I_{2\pi}$} \label{Fig:TdepV}
\end{figure*}

When $I_S(\varphi)=I_{2\pi}\sin\varphi$, the equation can be analytically solved\cite{}. For $I>I_{2\pi}$, the voltage oscillates in time with a period $T_{2\pi}$, and the dc voltage $\langle V(t)\rangle$ is as expected proportional to the frequency of the oscillations $f_{2\pi}=1/T_{2\pi}$:
\begin{eqnarray}
T_{2\pi}&=&\frac{\hbar}{2eR_n}\int_0^{2\pi} \frac{d\varphi}{I-I_{2\pi}\sin(\varphi)}=\frac{h}{2e}\frac{1}{R_n\sqrt{I^2-I_{2\pi}^2}},\\
\langle V\rangle&=&\frac{h}{2eT_{2\pi}}=R_n \sqrt{I^2-I_{2\pi}^2}.
\end{eqnarray}
For bias currents slightly exceeding the critical current, $I\simeq I_{2\pi}$, the time-dependent voltage $V(t)$ is highly non-sinusoidal (see Fig.\ref{Fig:TdepV}a), as a result of the non-linearity of the RSJ equation. This plays an important role in the case of a $4\pi$-periodic contribution to the supercurrent, as discussed below. In contrast, the response is almost harmonic for higher biases, and $V(t)$ becomes sinusoidal for $I\gg I_{2\pi}$.

\paragraph{Case of an additional $4\pi$-periodic contribution to the supercurrent}

Following the same reasoning as for $2\pi$-periodic supercurrent, we can define $f_{4\pi} =2/T_{4\pi}$ with the period defined by the following integral"
\begin{eqnarray}
T_{4\pi}&=&\frac{\hbar}{2eR_n}\int_0^{4\pi}\frac{d\varphi}{I-I_{2\pi}\sin(\varphi)-I_{4\pi}\sin(\varphi/2)}=T_1+T_2,\\
T_{1}&=&\frac{\hbar}{2eR_n}\int_0^{2\pi}\frac{d\varphi}{I-I_{2\pi}\sin(\varphi)-I_{4\pi}\sin(\varphi/2)},\\
T_{2}&=&\frac{\hbar}{2eR_n}\int_{2\pi}^{4\pi}\frac{d\varphi}{I-I_{2\pi}\sin(\varphi)-I_{4\pi}\sin(\varphi/2)}.
\label{eq.period4pi}
\end{eqnarray}
Since the term $I_{4\pi}\sin(\varphi/2)$ has contributions of opposite signs in $T_1$ and $T_2$, these two time scales can be very different (in particular when $I\sim I_c+I_{4\pi}$), thus affecting the periodicity of the dynamics.
When the current $I$ slightly exceeds the critical current, such that $I\sim I_c+I_{4\pi}$, one observes that $T_1\gg T_2$, and the system exhibits a non-sinusoidal $4\pi$-periodic behavior (Fig.\ref{Fig:TdepV}b). In contrast, for high currents ($I\gg I_c+I_{4\pi}$), $T_1\simeq T_2$ and therefore $f_{4\pi} \simeq f_{2\pi}$. The junction exhibits a dominating $2\pi$-periodic component, with weaker remaining $4\pi$-periodic modulations. This crossover originates in the highly non-linear and non-harmonic response of a Josephson junction for currents just above $I_c$. 

These observations are a natural behavior of the RSJ model as long as the CPR has both $2\pi$- and $4\pi$-periodic contributions. We believe that they are central to understand our experimental results. They explain in particular why the $4\pi$-periodic contribution can be optimally observed at low voltages. This model has already provided an explanation for the behavior of topological Josephson junctions when an additional rf current is applied (Shapiro response, see the Supplementary Information of Refs.\cite{Wiedenmann2016, Bocquillon2016} for detailed discussions).We show below that this reasoning, once adapted to the shunt circuit used in the present experiment, is likely to explain our experimental observations on the Josephson emission.

\begin{figure*}[h!]
\centerline{\includegraphics[width=\textwidth]{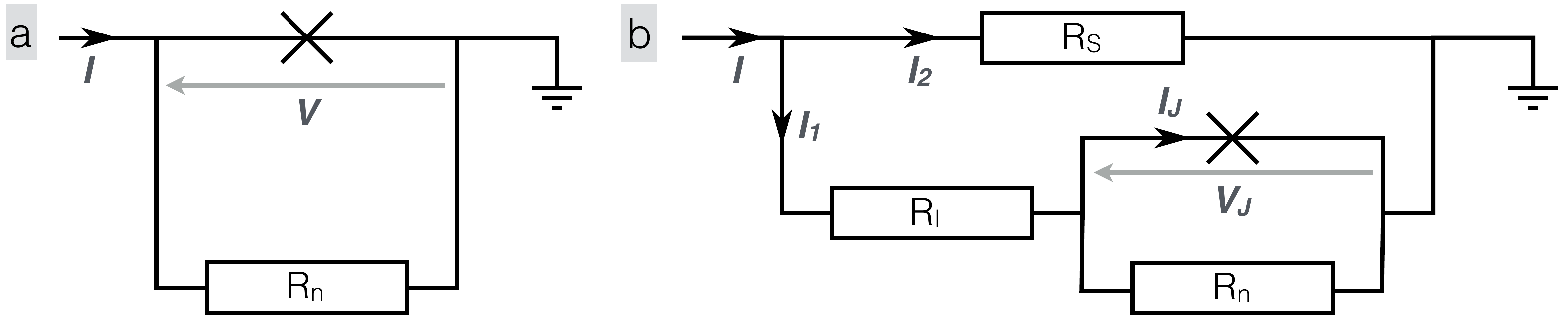}}
\caption{{\bf Circuits in RSJ models -} a) Standard RSJ framework, with the Josephson junction carrying a supercurrent $I_S(\varphi)$ in parallel with a shunt resistance $R_n$. b) Modified RSJ circuit taking into account additional measurement setup with resistors $R_I$ and $R_S$.} \label{Fig:SOMRSJsetups}
\end{figure*}

\subsection{Extended RSJ model with shunt circuit}

We first need to derive and adapt the equations of motion to the experimental shunt circuit, shown in Fig.\ref{Fig:SOMRSJsetups}b. The junction is as before represented by its CPR $I_S(\varphi)$ and its resistance $R_n$, but resistors $R_I$ and $R_S$ are added. Applying Kirchhoff laws, one extracts the modified equations of motion for the experimental setup, that read:
\begin{eqnarray}
I&=&I_S(\varphi) \left(1+\frac{R_I}{R_S}\right)+\frac{\hbar}{2e\tilde{R}_n}\frac{d\varphi}{dt}
\label{Eq:RSJmod}
\end{eqnarray}
where 
\begin{eqnarray}
\frac{1}{\tilde{R}_n}&=&\left(\frac{1}{R_n}+\frac{1}{R_S}+\frac{R_I}{R_S R_n}\right).
\end{eqnarray}
One sees that Eq.(\ref{Eq:RSJmod}) is identical to the standard RSJ equation (Eq.(\ref{Eq:RSJ})) with substitutions $R_n\to \tilde R_n$ and $I_c=\max_\varphi I_S(\varphi)\to I_c  \left(1+\frac{R_I}{R_S}\right)$. Simulations performed in the standard RSJ model can be readily adapted to this new setup. Besides, the experimental data is more naturally presented as a function of $I_1$ rather than $I$, which is obtained from $I_1=\frac{I-\frac{V_J}{R_S}}{1+\frac{R_I}{R_S}}$.

\subsection{Simulations of Josephson emission}

To compare the simulations with our experimental data, it is more convenient to set $I_c$ directly (rather than  $I_{2\pi}$ and $I_{4\pi}$ independently). We use the following parametrization :
\begin{eqnarray}
x&=&\frac{\ I_{4\pi}}{I_{2\pi}},\quad I_{2\pi}=\frac{32}{\left(3x+\sqrt{32+x^2}\right)\sqrt{32-2(x^2-\sqrt{32x^2+x^4})}}I_c.
\label{ratios}
\end{eqnarray}
With these notations, the critical current is set to $I_c$ and the ratio of $4\pi$- to $2\pi$-periodic supercurrents is tuned by $x$. 

\begin{figure*}[h!]
\centerline{\includegraphics[width=0.8\textwidth]{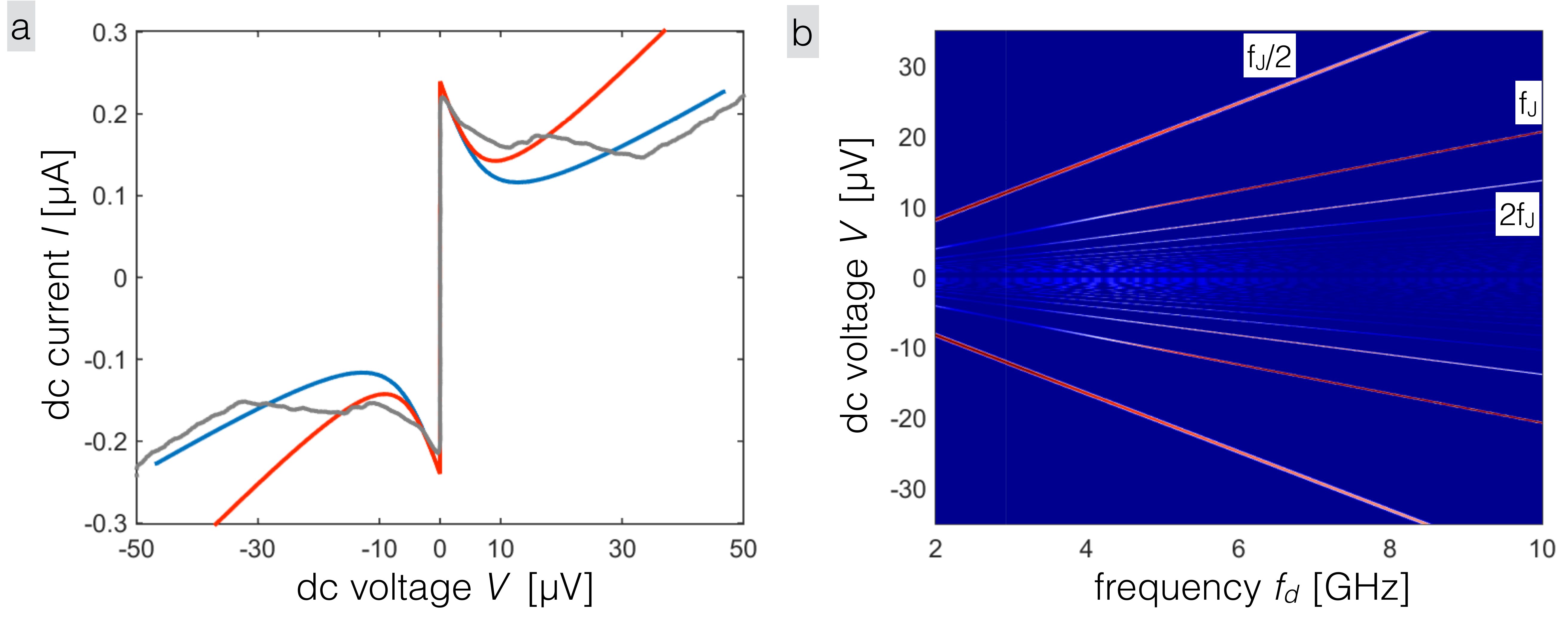}}
\caption{{\bf RSJ simulations -} a) Simulated $I$-$V$ curves compared with measured data at $V_g=\SI{-0.55}{\volt}$ (grey line). The simulations are performed for $R_I=R_S=\SI{24}{\ohm},\, R_n=\SI{130}{\ohm},\, I_{4\pi}=\SI{100}{\nano\ampere},\,I_c=\SI{240}{\nano\ampere}$ and $x=0.6$ (red line); and $R_I=R_S=\SI{24}{\ohm},\, R_n=\SI{220}{\ohm},\, I_{4\pi}=\SI{100}{\nano\ampere},\, I_c=\SI{240}{\nano\ampere}$ and $x=0.5$. b) Simulated Fourier transform of the voltage $V$ in the junction, as function of detection frequency $f_d$ and voltage $V$, for the simulation parameters of the blue line (higher resistance). A good qualitative agreement with Fig.2e can also be found with these parameters.} \label{Fig:SOMSimus}
\end{figure*}

\paragraph{Fitting the $I$-$V$ curve} We first optimize the fitting of the $I$-$V$ curve (experimental data of Fig.2b of the main text) by choosing the value of $I_c$ and $R_n$. $R_I$ and $R_S$ are set to the value used in the setup, i.e. $R_I=R_S=\SI{25}{\ohm}$. We observe that $x$ has a marginal effect on the $I$-$V$ characteristic, and can be ignored in this part. As depicted in Fig.\ref{Fig:SOMSimus}, we find the best agreement for $I_c\simeq \SI[separate-uncertainty]{240(10)}{\nano\ampere}$ and for two different values of $R_n$. For $R_n\simeq \SI[separate-uncertainty]{130(15)}{\ohm}$ (presented in the main text, here as a red line), we obtain a good agreement for low voltages only.  For $R_n\simeq \SI[separate-uncertainty]{220(15)}{\ohm}$ (blue line) the agreement is not as good for low voltages, but remains decent also for higher voltages.

\paragraph{Estimating the $4\pi$-periodic supercurrent} The last parameter that needs to be evaluated is the fraction $x$ of $4\pi$-periodic supercurrent. From the RSJ simulations of $V(t)$, we compute the spectral density of noise $|V(f)|$ by fast Fourier transform. As explained previously, the Josephson emission mostly occurs at $f_{\rm J}/2$ at low voltages up to a crossover voltage $V_{4\pi}$ that increases with $x$. In the experimental data, $V_{4\pi}\simeq \SI{12}{\micro\volt}$. In our simulations, the crossover is not as abrupt, but reaches this value for $x\simeq\num[separate-uncertainty]{0.6(1)}$ (similar for both values of $R_n$), so that $I_{4\pi} \simeq 0.4 I_c$. From these simulations, we find the estimate $I_{4\pi}\simeq80-\SI{120}{\nano\ampere}$ which exceeds the expected contribution of two edge modes ($\lesssim\SI{50}{\nano\ampere}$). However, though the physics previously described is quite universal, this estimate of $I_{4\pi}$ has to be taken cautiously as it is more strongly dependent on the model and choice of parameters.

\subsection{Inductive effects and period doubling}

Our experimental setup comprises bond wires between the resistors and the device chip. Bond wires add an inductive contribution to the shunt circuit, estimated around a few \si{\nano\henry} (typically $\SI{1}{\nano\henry\per\milli\meter}$), Inductance in shunt circuits strongly modify the RSJ equation (by adding higher order derivatives) and lead to complex dynamical behavior of Josephson junctions\cite{Sullivan1970,Miracky1983,Whan1995,Cawthorne1998}. Of particular interest, it can lead to period doubling and could be responsible for a Josephson radiation at $f_ {\rm J}/2$. 
Several observations in our experiments show that this explanation is in fact unlikely.

First, the presence of these phenomena is governed by the dimensionless inductance and capacitance parameters in the shunt branch\cite{Cawthorne1998}, $\beta_L=\frac{2eI_cL}{\hbar},\beta_C=\frac{2eR_S^2 I_c C}{\hbar}$, where $L$ and $C$ are respectively the inductance in the shunt branch and the capacitance of the junction. We find that, indeed, the inductive parameter is rather large (around $\beta_L\simeq5$ for the parameters at $V_g=\SI{-0.55}{\volt}$). However the shunt resistance and the capacitance of the junction are small and yield $\beta_C\simeq 0.01$. In these conditions, it has been predicted\cite{Whan1995,Cawthorne1998} that inductive effects have limited consequence, and the junction should not be subject to non-linear subharmonic oscillations. However the threshold value are probably model-dependent and should be taken with care.

Second, this period doubling is usually accompanied by many other complex patterns such as transition to chaos, relaxation oscillations, long transients\cite{}. The exact dynamical behavior of the junction thus becomes extremely sensitive to the junctions parameters (bias current, capacitance and inductance, etc.). In particular, emission at $f_ {\rm J}/3$, $f_ {\rm J}/4$, etc. would likely accompany Josephson radiation at $f_ {\rm J}/2$. Despite massive changes in $I_c$, we do not detect any emission at any other subharmonic of the Josephson frequency than $f_ {\rm J}/2$. Typical signatures of chaotic behavior\cite{Whan1995} are not observed either in the $I$-$V$ curves.

Finally, both the topological and trivial weak links have been measured in the same setup, up to minor differences in the bonding schemes. As both devices have similar design (hence similar parallel capacitance) and similar critical currents, inductively driven subharmonic oscillations would likely be observed in both systems. However a thorough study of the trivial weak link has excluded the presence of emission at $f_{\rm J}/2$ or any other subharmonic in this system.

\section{Electromagnetic environment and influence on Josephson emission}

In this section, we provide measurements aiming at characterizing the electromagnetic environment of the Josephson junction, and correlate different sets of features with the differential conductance of the device and with emission features observed at high frequencies.

\subsection{Characterization of the environment}

The bonding of the Josephson junction to a coaxial line yields a very simple way to measure Josephson emission over a wide range of frequency, as illustrated by our measurements. However, the \SI{50}{\ohm} coaxial line and the junction are not impedance matched. Impedance matching is in general very hard to achieve in a broad band for very different and varying impedances as such as here. As a consequence, resonances inevitably appear in the coupling of the device to the amplification scheme, that can in turn modify the response of the device.

\begin{figure*}[h!]
\centerline{\includegraphics[width=0.8\textwidth]{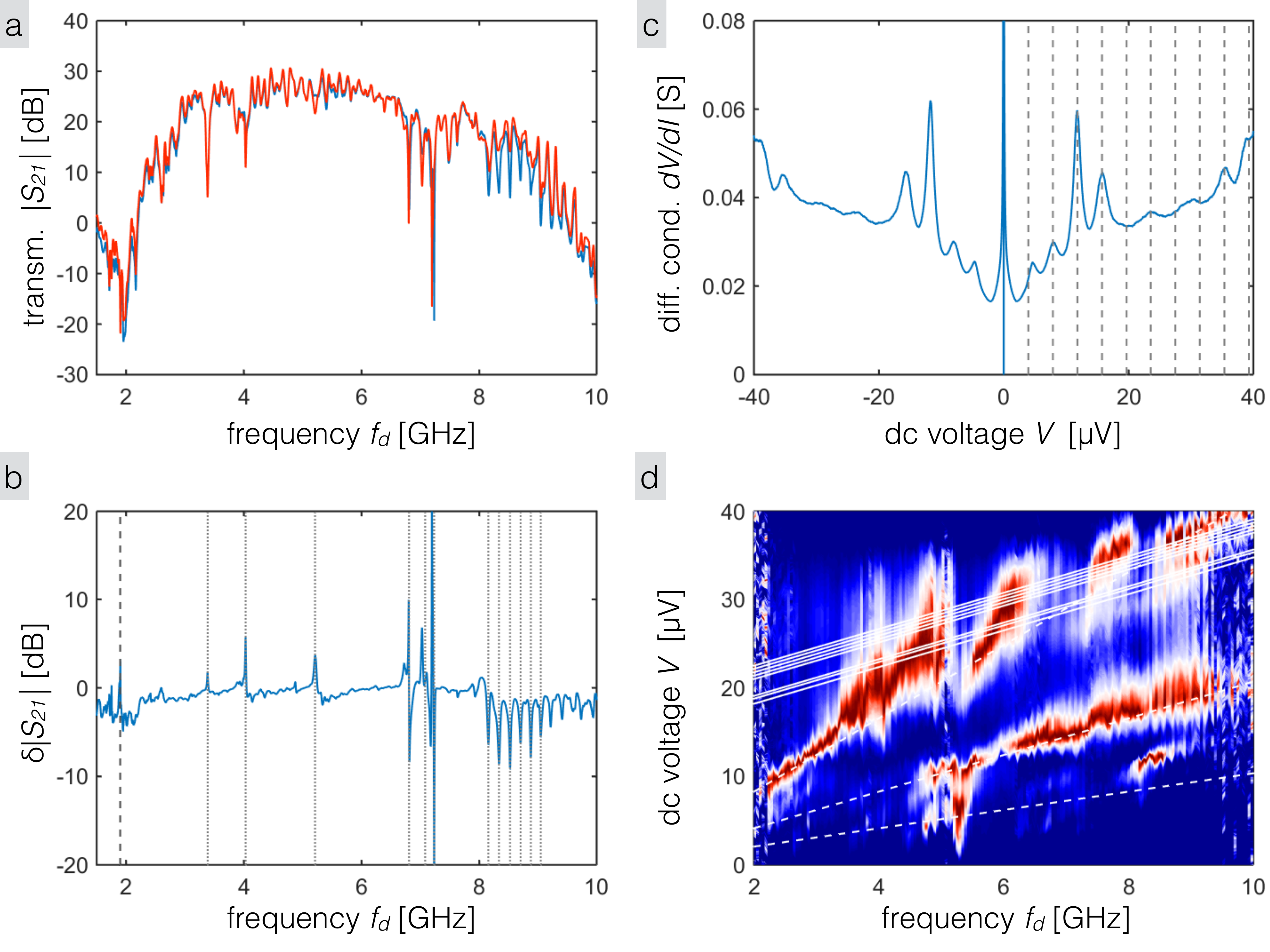}}
\caption{{\bf Comparison of Reflection measurements with transport and emission -} a) Transmission magnitude $|S_{21}|$ as function frequency measured using a vector network analyzer. b) Difference in transmission magnitude $\delta |S_{21}|$ as a function of frequency. Dashed lines indicate features which are sensitive to changes in the junction impedance. c.) Plot of junction differential conductance measured at $V_{g}=\SI{0}{\volt}$. Dashed lines indicate self-induced Shapiro steps  corresponding to the resonance at $f_d=\SI{1.9}{\giga\hertz}$. d.) Two-dimensional plot of the power emitted, as function of voltage $V$ and detection frequency $f_d$, measured at $V_{g}=\SI{0}{\volt}$. For better visibility, the data is normalized to its maximum for each frequency. Solid lines indicate replicas of the main emission line offset by the resonances observed in the reflection measurements shown in (a). Dashed lines indicate the expected resonance lines at $2f_{\rm J}$ (lower) and $f_{\rm J}$ (upper).} \label{Fig:S2Resonances}
\end{figure*}

Using the directional coupler, it is possible to send an rf excitation (through port 1) towards the sample and measure how it is transmitted/reflected to the amplification line (port 2). In Fig.\ref{Fig:S2Resonances}a, we measure the transmission $S_{21}$ as function of frequency with a vector network analyzer. As expected the signal drops outside the range $f_d=2-\SI{10}{\giga\hertz}$ corresponding to bandwidth of the rf cryogenic amplifier. In between, many resonances are visible. These resonances can occur due to standing waves between rf components (amplifiers, circulators, connectors, etc.) in both the excitation and amplification lines, and do not depend on the device under study. To isolate the ones that are related to the coupling of the device, we measure two sets for different dc excitations $I_{\rm exc}$ on the junction (Fig.\ref{Fig:S2Resonances}a). The modification of the impedance of the junction results in modifications of the spectrum. Some resonances have a different amplitude, or undergo a small frequency shift. The difference between both spectrum $\delta |S_{21}|$ is presented in Fig.\ref{Fig:S2Resonances}b and reflects these modifications, enabling the identification of the features related to the Josephson junction itself (dashed and dotted grey lines), some of these resonances exhibiting quality factors of several hundreds.

\subsection{Possible influence on emission measurements}
\paragraph{Self-induced steps}
The first consequence of unintentional resonant structures near the device is the appearance of Shapiro steps in the device : at resonance, the emitted Josephson radiation is fed back onto the junction resulting in self-induced Shapiro steps. The differential resistance of our device $dV/dI_{\rm exc}$ exhibits a series of equidistant peaks. This appear to be in very good agreement with the Shapiro steps associated to the resonance at $f_d=\SI{1.9}{\giga\hertz}$ (dashed line in Fig.\ref{Fig:S2Resonances}b). Though the differential resistance can modify the amplitude of the emitted radiation \cite{Likharev1986}, no strong connection to our measurements of the power has been established.
\paragraph{Josephson emission}

Second, one could expect that resonances influence the conditions in which the Josephson emission is radiated. We believe our measurements exhibit replicas of the Josephson emission, possibly due to resonances in the coupling, namely emission features at $f_{\rm J}+f_r$ where $f_r$ is a resonance frequency in the electromagnetic environment of the Josephson junction. These features are illustrated on Fig.\ref{Fig:S2Resonances}d, by placing white lines at $f_{\rm J}+f_r$ for every value of $f_r$ reported as a dotted grey line on Fig.\ref{Fig:S2Resonances}b.

Similar effects have been theoretically and experimentally investigated in the regime of dynamical Coulomb blockade \cite{Hofheinz2011,Leppakangas2014}. When the junction is embedded in a cavity resonating at $f_r$, two-photon processes can give rise to replicas of the main emission line (at $f_{\rm J}$) shifted by the energy of a photon, namely $f_{\rm J}+f_r$. Given the impedance of our device $R_s, R_n\ll R_{\rm K}=\frac{h}{e^2}$, our device is not in the appropriate regime to observe dynamical Coulomb blockade effects. In particular, two-photon processes are second-order in $R_n/R_{\rm K}$ and should always be much less visible as standard emission at $f_J$. As such they cannot solely explain the observation of radiation at $f_{\rm J}/2$. Besides, when no radiation is detected at $f_{\rm J}/2$, the complex pattern of Fig.\ref{Fig:S2Resonances}d is absent or barely visible (see Fig.2d and f in the main text). However, we speculate that the interplay of resonant two-photon processes and anomalous emission at $f_{\rm J}/2$ are a possible explanation for the observed high-frequency features.

Alternatively, embedding of a Josephson junction described by RSJ equations into a resonator has been theoretically investigated in Refs.\cite{Gubankov1976,Larsen1991}. Self-induced Shapiro steps are then predicted to occur, as well as radiation at the resonator frequency $f_r$, but no explicit prediction has been made concerning emission at $f_{\rm J}+f_r$ and further investigation is required.

\section{Additional experimental results}

\subsection{Gate voltage dependence of the non-topological weak link}

\begin{figure*}[h!]
\centerline{\includegraphics[width=0.8\textwidth]{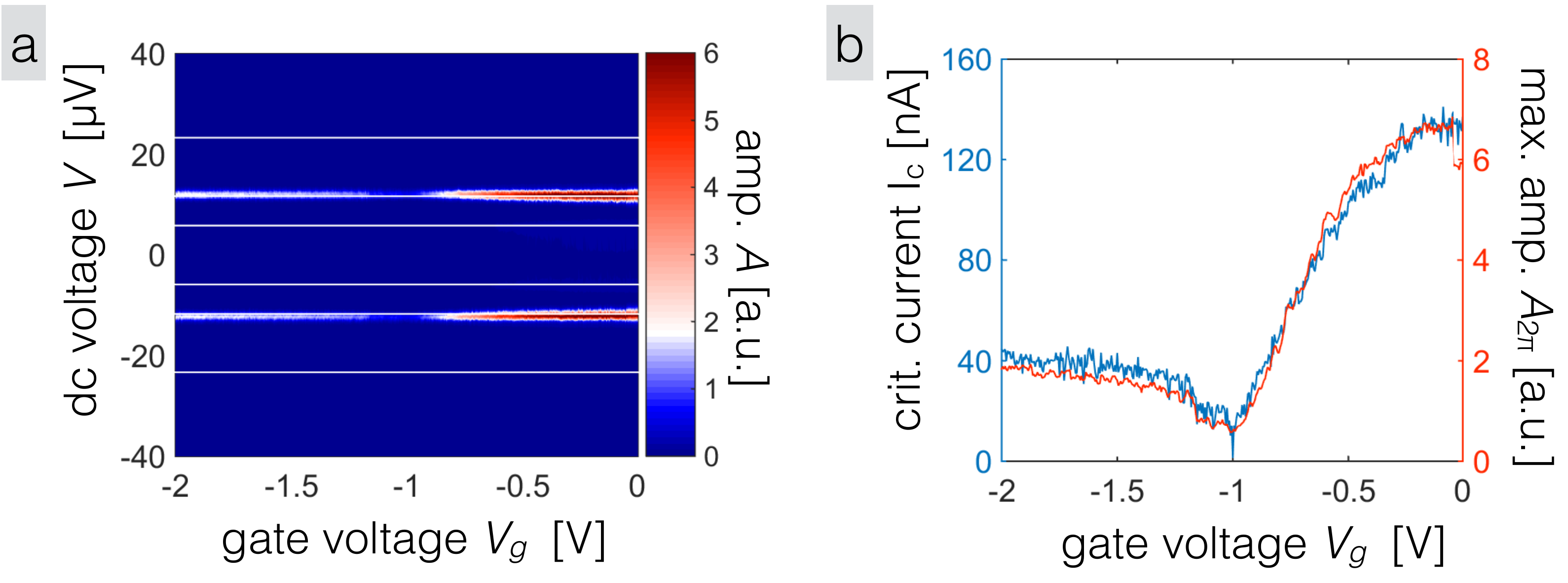}}
\caption{{\bf Gate dependence of the non-topological weak-link -} a) 2D map of the collected rf amplitude $A$ as function of bias voltage $V$ and gate voltage $V_g$, for a detection frequency $f_d=\SI{5.64}{\giga\hertz}$. b) Extracted peak rf amplitude $A_{2\pi}$ and dc critical current $I_c$ vs gate voltage $V_g$.} \label{Fig:S1GateDepRadiance}
\end{figure*}

In this section, we show that the non-topological weak link follows to a good accuracy the expected behavior $A\propto I_c$ between the collected rf amplitude $A$ and the critical current $I_c$ \cite{Likharev1986}. The latter is tuned via the gate voltage $V_g$. In Fig.\ref{Fig:S1GateDepRadiance}a, we present as a colormap the amplitude $A$ as function of voltage $V$ and gate voltage $V_g$. The amplitude of the measured emission line at $f_{\rm J}$ scales exactly with the amplitude of the dc supercurrent, as shown in Fig.\ref{Fig:S1GateDepRadiance}b. Moreover, it is additionally shown in the 2D map that radiation at $f_{\rm J}/2$ is completely absent in this device. 

\subsection{Emission in HgTe-based 3D topological insulators}

\begin{figure*}[h!]
\centerline{\includegraphics[width=0.8\textwidth]{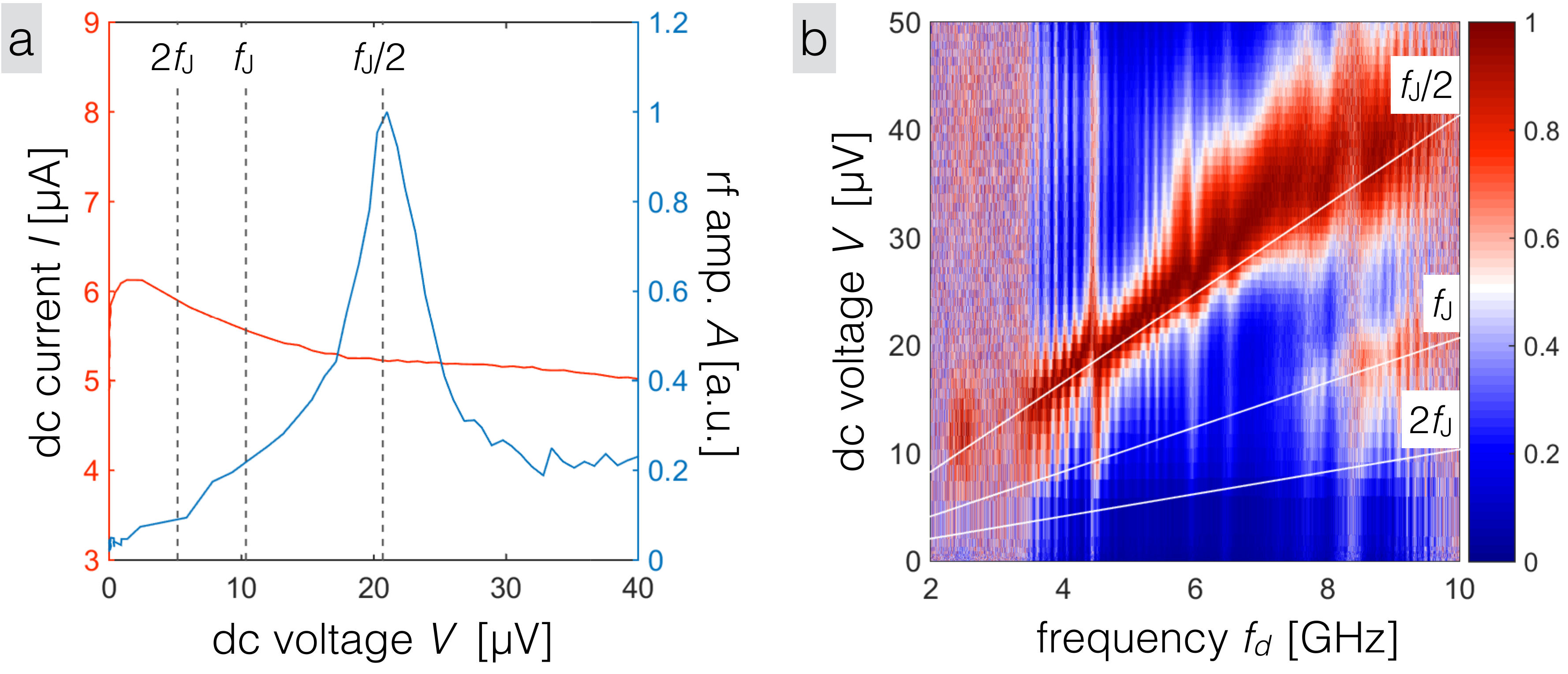}}
\caption{{\bf Emission at $f_{\rm J}/2$ in HgTe-based 3D topological insulators -} a) $I$-$V$ curve (red line) and normalized emission magnitude $A$ (blue line) for a 3D topological insulator weak link. The radiation is collected at fixed detection frequency $f_d=\SI{5}{\giga\hertz}$. b) 2D map of the collected rf amplitude $A$ as function of bias voltage $V$ and detection frequency $f_d$. For better visibility, the data is normalized to its maximum for each frequency.} \label{Fig:S3Rock3D}
\end{figure*}

In this section, we show measurements of the Josephson emission on weak links made of \SI{70}{\nano\meter}-thick strained layers of HgTe. Such layers have been demonstrated to be 3D topological insulators \cite{Bruene2011,Bruene2014}. In a previous work \cite{Wiedenmann2016}, we have detected the presence of a weak $4\pi$-periodic supercurrent flowing in such Josephson junctions, as a signature of a single topologically protected Andreev doublet in agreement with theoretical predictions \cite{Tkachov2013}. The measurement of anomalous emission at $f_{\rm J}/2$ confirms the presence of this $4\pi$-periodic supercurrent.

Though the shunt circuit is similar to that of Fig.1b (main text), the measurements have been performed slightly differently. Here the detection frequency $f_d$ is swept at a fixed value of the bias current. The disadvantages of this method are an increased time consumption and the difficulty to correct for slow drifts of the background. The measurements are shown in Fig.\ref{Fig:S3Rock3D}. In Fig.\ref{Fig:S3Rock3D}a, the $I$-$V$ curve of the device is presented together with the normalized rf amplitude $A$ at a detection frequency $f_d=\SI{5}{\giga\hertz}$. As previously, a clear peak is observed at $f_{\rm J}/2$.
In Fig.\ref{Fig:S3Rock3D}b, the 2D map of the normalized amplitude $A$ as function of voltage $V$ and detection frequency $f_d$ is shown. As in the main text, the $4\pi$-periodic supercurrent is seen to dominate the low-frequency/voltage regime, while the conventional $2\pi$-periodic supercurrent is recovered at $f_d=\SI{7.5}{\giga\hertz}$.
In this plot with high frequency resolution, it is seen that the frequency of the radiated signal is slightly shifted from the expected value of $f_{\rm J}/2$. This shift seems to follow oscillations in the transmitted signal and could reflect again the influence of resonances in the surrounding electromagnetic environment, as discussed in Section II. The study of line widths yields here $\delta V\simeq2-\SI{8}{\micro\volt}$ at half-width, corresponding to $\tau_{4\pi}\simeq 0.25-\SI{1}{\nano\second}$.

\bibliographystyle{unsrt}

\bibliography{BibJosephsonEmission.bib}

\end{document}